\documentclass[twocolumn,prx,superscriptaddress,aps]{revtex4}
\usepackage{amsmath}
\usepackage{graphicx}
\usepackage[squaren]{SIunits}
\usepackage{sistyle}
\usepackage{hyperref}

\usepackage{color}
\definecolor{mygreen}{rgb}{0,0.5,0} 
\definecolor{mygrey}{rgb}{0.5,0.5,0.5} 
\definecolor{myred}{rgb}{0.75,0,0} 
\definecolor{myblue}{rgb}{0,0,0.75} 
\definecolor{mymagenta}{cmyk}{0,1,0,0.12} 
\definecolor{mycyan}{cmyk}{1,0,0,0.12} 
\definecolor{myorange}{rgb}{1,0.5,0} 
\definecolor{myviolet}{rgb}{0.5,0.3,1} 

\usepackage{ulem}

\begin{document}

\newcommand{\mytitle}{Maltese cross coupling to individual cold atoms in free space}

\title{\mytitle}

\newcommand{\ICFO}{ICFO - Institut de Ciencies Fotoniques, The Barcelona Institute of Science and Technology, 08860 Castelldefels, Barcelona, Spain}
\newcommand{\ICREA}{ICREA - Instituci\'{o} Catalana de Recerca i Estudis Avan{\c{c}}ats, 08010 Barcelona, Spain}
\newcommand{\ZH}{State Key Laboratory of Modern Optical Instrumentation, College of Optical Science and Engineering, Zhejiang University, Hangzhou 310027,China}

\author{Natalia Bruno}
\affiliation{\ICFO}
\author{Lorena C. Bianchet}
\affiliation{\ICFO}
\author{Nan Li}
\affiliation{\ZH}
\author{Vindhiya Prakash}
\affiliation{\ICFO}
\author{Nat\'{a}lia Alves}
\affiliation{\ICFO}
\author{Morgan W. Mitchell}
\affiliation{\ICFO}
\affiliation{\ICREA}

\begin{abstract}
We report on the simultaneous observation from four directions of the fluorescence of single $^{87}$Rb atoms trapped at the common focus of four high numerical aperture (${\rm NA}=0.5$) aspheric lenses. We use an interferometrically-guided  pick-and-place technique to precisely and stably position the lenses along the four cardinal directions with their foci at a single central point. The geometry gives right angle access to a single quantum emitter, and will enable new trapping, excitation, and collection methods. The fluorescence signals indicate both sub-Poissonian atom number statistics and photon anti-bunching, showing suitability for cold atom quantum optics.   
\end{abstract}

\maketitle

\begin{figure*}[ht]
\centering
\includegraphics[width=0.7\columnwidth]{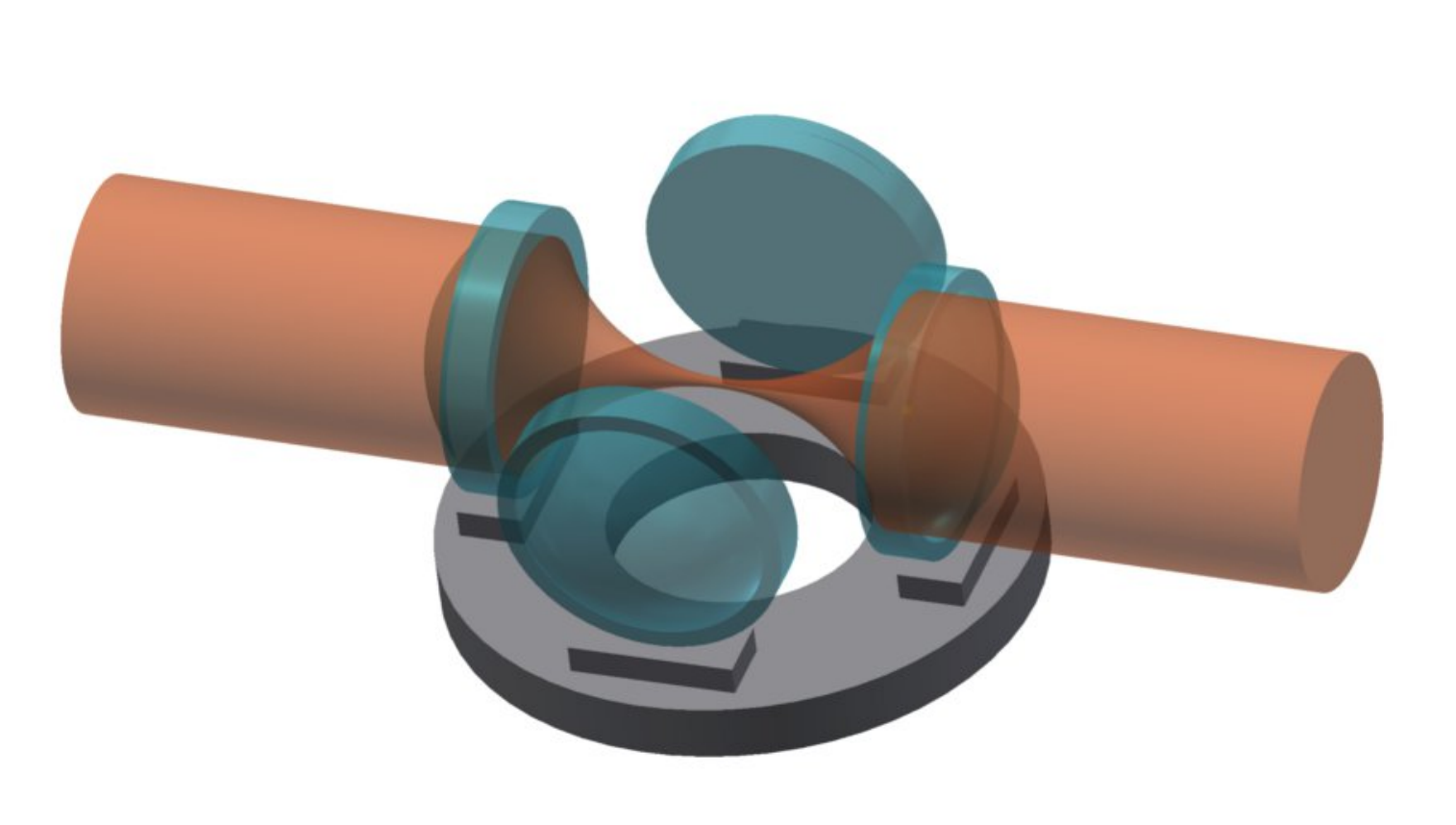}
\hspace{5mm} \includegraphics[width=0.4\columnwidth]{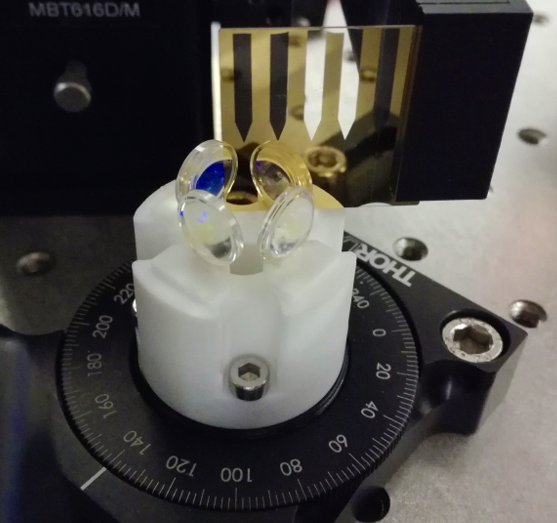}
 \includegraphics[width=0.5\columnwidth]{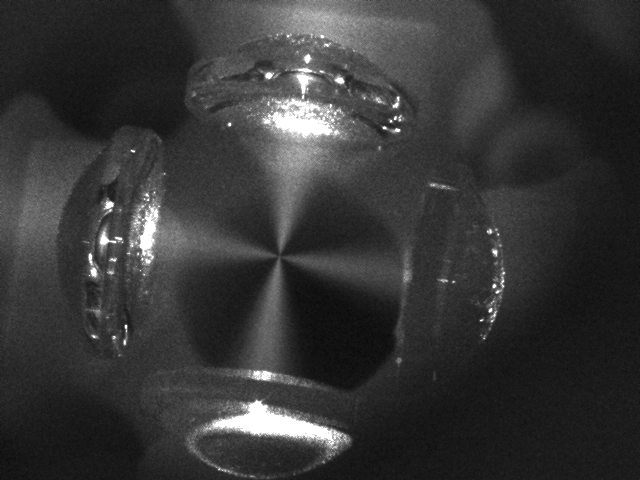}
\includegraphics[width=0.9\columnwidth]{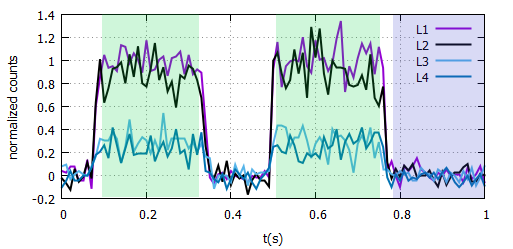}
\hspace{0mm} \includegraphics[width=0.9\columnwidth]{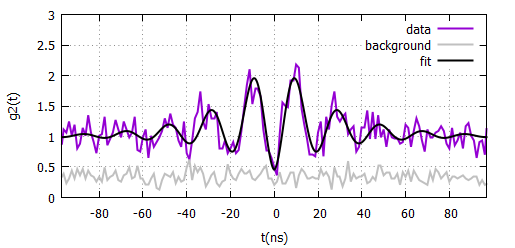}
\caption{Four-lens geometry and atomic signals.  Top left: illustration of central optical components and support.  Four lenses (cyan) are positioned to minimize aberrations and affixed to a Macor ceramic substrate (grey) with ultra-high-vacuum compatible epoxy. The four lenses share a single focus which lies within their diffraction-limited field of view.  A single-beam FORT (orange) is used to trap a single atom from a co-located MOT (not shown) and hold it at the common focus. Top center: The four lenses glued in place and being tested by placing a gold first-surface mirror with a transmissive aperture centered at their mutual focus. Top right: an intermediate step of the alignment in vacuum, using \SI{780}{\nano\meter} light to excite resonance fluorescence of a $^{87}$Rb vapour to visualize the overlap of the foci. Bottom left: fluorescence signals acquired into single-mode fibre from each of the four lenses for periods with one atom (green shading) or with no atom (white and blue shading).  In agreement with theory, lenses L1 and L2 (along the FORT axis) show higher collection efficiency than right-angle lenses L3 and L4, due to the elongated shape of the atom's spatial  distribution at finite temperature.  A background, due to scattering of the MOT beams, of about 10 \%  full scale has been subtracted.  Bottom right:  normalized second-order autocorrelation function showing anti-bunching, confirming the sub-Poissonian atom number.  }
\label{fig:Overview}
\end{figure*}

\section{Introduction}

\newcommand{\NA}{{\rm NA}}

Optically trapped neutral atoms are an important platform for quantum technologies and studies of fundamental quantum optics.  In strongly-focused optical traps the atom number exhibits strongly sub-Poissonian statistics \cite{SchlosserN2001}, facilitating the isolation of individual atoms and organization into 1D \cite{BernienN2017} 2D \cite{LabuhnN2016, PerczelPRL2017} and even 3D arrangements \cite{BarredoN2018}. When multiple traps are tunnel-coupled, multi-atom interferences \cite{KaufmanS2014} resembling the Hong-Ou-Mandel effect can be observed.  Long-range atom-atom interactions can be produced using Rydberg states \cite{GaetanNP2009, UrbanNP2009, JakschPRL2000}, and are being actively pursued for application in quantum simulation \cite{LabuhnN2016, BernienN2017} and quantum computation \cite{SaffmanJPB2016}.  Short-range dipole-dipole interactions are predicted to dramatically modify the optical properties of matter in sub-wavelength arrays of single atoms \cite{AsenjoGarciaPRX2017}. These processes can be studied at the single-quantum level because high numerical aperture optics enable strong single-atom/single-photon interactions in free-space \cite{TeyNP2008, AljunidPRL2009}, without optical cavities. Many effects of current interest are not compatible with cavity enhancement because they intrinsically concern the spatial  \cite{AsenjoGarciaPRX2017, PerczelPRL2017} and temporal \cite{AljunidPRL2013, GuerinPRL2016} behaviour of  propagating fields.

All of the above applications require high-NA, multi-wavelength access to a small region in which the atoms are trapped. Similar techniques have also been applied to individual trapped ions \cite{MaiwaldNP2009, SlodickaPRL2010, SondermannJMO2013}. Initial experiments used a single high-NA lens or objective \cite{SchlosserN2001}, and soon thereafter developed co-linear pairs of high-NA lenses for bi-directional access \cite{BeugnonNP2007, TuchendlerPRA2008}. Already the use of lens pairs offers an important advantage in coupling strength and allows the observation of nonlinearities at the single atom \cite{ChinNC2017}.  Achieving a still greater coupling is a major challenge, and has motivated exotic optical techniques \cite{SondermannJMO2013}. Here we describe an approach using four high-NA aspheric lenses in vacuum. We use interferometric methods to precisely position and align the four lenses. This approach achieves simultaneous, diffraction-limited performance at $\NA = 0.5$ for wavelengths \SI{780}{\nano\meter}, \SI{795}{\nano\meter} and \SI{852}{\nano\meter}, enabling strong interaction with the D$_1$ and D$_2$ lines of atomic rubidium, plus strongly-focused optical dipole trapping. We confirm the diffraction-limited performance with fluorescence measurements on a single atom trapped at the mutual focus of the four lenses.  

Relevant figures of merit for coupling to single atoms are the probability with which a single atom can be excited by a single photon \cite{StrobinskaEPL2009}, scatter a photon out of a beam \cite{TeyNJP2009}, and emit a photon into a defined spatial mode. These three measures are related in that they are all functions of the overlap of an optically-defined mode with the dipole radiation pattern of the atom.  Chin et al. \cite{ChinNC2017} have shown that beams from multiple directions can be coherently combined to enhance these figures of merit beyond what is possible with a single input beam.  Here we consider collection of a single photon emitted by the atom into a mode formed as the superposition of $N$ gaussian beams, one behind each of $N$ aspheres.  The net collection efficiency is $\eta_{N} = N \eta_1$, where $\eta_1$ is the collection efficiency into a single gaussian mode behind a single asphere.  For our beam waist of \SI{2.2}{\milli\meter}, full ${\rm NA} = 0.5$, and focal length $f = \SI{8}{\milli\meter}$ we compute \cite{inprep} $\eta_1 = 0.049$ and thus $\eta_4 =  0.194$. This can be compared against prior work \cite{ChinNC2017} with two ${\rm NA} = 0.75$ aspheres of focal length $f = \SI{5.95}{\milli\meter}$ and beam waist \SI{2.7}{\milli\meter}, for which $\eta_1 = 0.11$ and thus $\eta_2 = 0.22$. Hence, we estimate a collection efficiency comparable to the state of the art record, with the advantage of having doubled the accessible directions and without the need of increasing the numerical aperture.

Right-angle access also enables new trapping geometries with no ``soft'' direction and smaller features in any direction. For example, a two-dimensional lattice can be implemented by using crossed standing waves at \SI{852}{\nano\meter}, with a lattice constant $d/ \lambda \approx 0.8 < 1$, necessary for observing the collective effects described in \cite{AsenjoGarciaPRX2017}. This will benefit studies of super- and sub-radiance \cite{WeissNJP2018}, selective-radiance \cite{AsenjoGarciaPRX2017}.
Furthermore, non-collinear collection of single photons scattered by individual atoms will improve the signal-to-noise ratios in experiments where the probing light is a source of background noise, as when investigating optical properties of dense atomic media \cite{JenneweinPRL2016}, time-reversal of spontaneous emission \cite{LeuchsPS2012, AljunidPRL2013}, and single-photon/single-atom interactions \cite{LeongNC2016}. In sum, the Maltese cross geometry both enables a new class of experiments, and improves the coupling efficiency, the key figure of merit for many free-space experiments.

As illustrated in Fig.~\ref{fig:Overview}, the optical system consists of four aspheric high-NA lenses in vacuum, affixed to a rigid ceramic support, with one lens along each of the cardinal directions. The lenses are positioned to nominally share the same focal point at the centre of the ceramic support, and can focus both the resonance wavelengths \SI{780}{\nano\meter} and \SI{795}{\nano\meter} used for laser cooling and spectroscopic manipulations, and also \SI{852}{\nano\meter}, a convenient wavelength for creation of structured conservative potentials, i.e. optical dipole traps and optical lattices. From Fig.~\ref{fig:Overview}, top right, one can see that of the four focused beams form a ``Maltese cross'' shape, which gives the configuration its name. One challenge of using single-element aspheres rather than multi-element lenses or objectives is the relatively small diffraction-limited field of view, and both precise positioning and tilt of the four lenses is critical to the strategy.  In what follows we describe in detail a solution to this alignment problem using a combination of interferometric techniques and micro-fabricated optical alignment aides. 

To test the system, we place it in ultra-high vacuum (UHV) with a source of $^{87}$Rb and magnetic field controls, to produce a magneto-optical trap (MOT) around the focus of the four lenses.  A beam at \SI{852}{\nano\meter} is then introduced through one of the high-NA lenses to create a wavelength-scale trapping region at the common focus, a tightly-focused far-off-resonance trap (FORT).  Due to light-assisted collisions, the trap can hold at most one atom, and switches randomly from a zero-atom to one-atom condition.   Fluorescence from the trapped atom is collected into single-mode fibres at the output of each of the four lenses.  Anti-bunching and Rabi oscillation are seen in the $g^{(2)}$ auto-correlation function of the collected fluorescence, confirming the single-atom occupancy of the trap. Observation of equal fluorescence signals from the two trap-axis lenses, and equal but weaker fluorescence signals from the two right-angle lenses, agrees with modeling of diffraction-limited collection from the prolate atomic probability distribution (long axis along the trap axis), that results from trapping in the single-beam FORT. Interference of light emitted in different directions can be achieved if the probability distribution is further compressed by a standing-wave \cite{ChinNC2017} or crossed-beam \cite{PalaciosNJP2018} FORT.

\section{Optical design}

\begin{figure}[ht]
\begin{center}
\includegraphics[width=1.0\columnwidth]{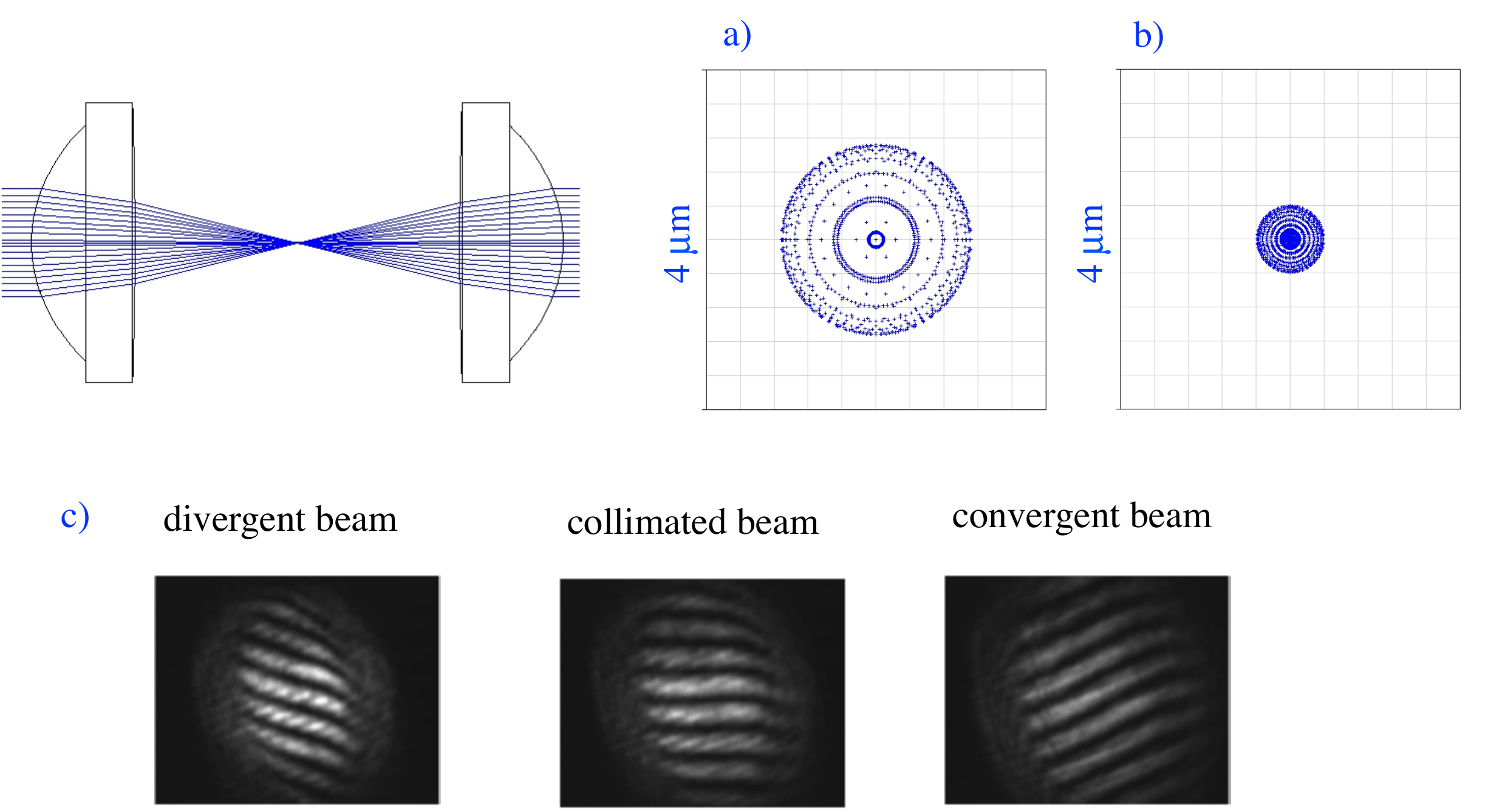}
\caption{ a) Spot diagram simulated in ZEMAX-EE at focus for a collimated beam,  spot size is $3\, \mu$m. b) Spot diagram at focus for a convergent beam (\SI{1.9}{m rad}), the spot size is $0.77\, \mu$m. c) SI fringes measured for a divergent ($\sim$\SI{1}{m rad}), collimated and convergent ($\sim$\SI{1}{m rad}) beam passing twice through the same aspheric lens, backreflected at focus by a gold mirror. The evident distortion in the first two images indicates the presence of spherical aberration.}\label{zemax}
\end{center}
\end{figure}

 We begin with calculations of expected optical performance, using an optical design program (ZEMAX-EE) and lens shape files supplied by the lens manufacturer.   The central elements in the design are the aspheric lenses.  After considering the commercially-available models, we selected one, Model 352240 from LightPath Technologies, ${\rm NA}=0.5$, that has already been used in similar experiments \cite{LucasBeguinThesisInstOpt2013}, and has proven to be diffraction limited over a wide spectral range and a relatively large field of view: \SI{\pm 25}{\micro \meter} in the transverse directions \cite{SortaisPRA2007} and \SI{\pm 47}{\micro\meter} in the longitudinal direction.
This aspheric lens, like most such lenses, is designed to be diffraction limited when used with a \SI{0.25}{m m}-thick glass laser window, whereas there is in fact no such window between the lens and atom in the foreseen trapping geometry. As such, this asphere is not initially diffraction limited in vacuum, but rather shows a small spherical aberration when used with a collimated input beam (here and throughout, we will describe, from a lens-centred perspective, the scenario of focusing light onto the atom, so that the ``input beam'' is approximately collimated while the ``output beam'' is strongly converging).    
Nonetheless, by changing the divergence of the input beam by \SI{1-2}{m rad},  one can introduce a wavefront error that compensates the spherical aberration, as shown in Fig.\ \ref{zemax}. 

Again with ZEMAX-EE, we simulate a linear focusing and collection system, composed of two achromatic doublets to prepare the input beam, the \SI{2}{\milli\meter}-thick glass window of the vacuum chamber and the two aspheric lenses, with all elements co-linear.  We quantify aberrations by the Strehl ratio $S$, which is the ratio of peak intensity in the image plane for the simulated optical system to peak intensity in the image plane for an ideal optical system with the same aperture and illumination, here a gaussian beam. The condition $S\geq0.8$ is a commonly taken to indicate diffraction-limited performance \cite{BornBook2013p468}.
We note that the beam-shaping doublets and window introduce negligible aberrations.

We estimate the waist and divergence of the beam at the first lens, the Strehl ratios after one ($S_1$) and two lenses ($S_2$), and the waist at focus such that the foci of the three wavelengths involved overlap, the values are shown in Table~\ref{tab1}.
These values represent a compromise between reducing aberrations and using the full numerical aperture of the aspheric lenses.
 
We use a wedged Shearing Interferometer ( SI, ThorLabs SI100) to measure and adjust the divergence of the beams.
We use the SI to adjust the displacement in the transversal plane and the tilt, as these misalignments cause wavefront errors that are detected by the SI as curvature of the observed fringes. 
 We find that the smallest beam-tilt angle $\theta$ for which we can detect a change in the tilt of the fringes is $\theta =$\SI{ 0.25}{\arcdeg}, which translates into a displacement of the focal point by \SI{27}{\micro \meter} and a decrease of the total Strehl ratio after a pair of lenses from 0.80 to 0.79 at \SI{795}{\nano \meter}. At $\theta = $\SI{ 0.5}{\arcdeg} we start seeing distortions of the fringes, and for this amount of beam-tilt we calculate $S = 0.74$. 
We can note that for higher wavelengths the fringes undergo less distortions and the Strehl ratio results in a higher value, however the amount of aberration calculated for the shortest wavelength involved in our system, \SI{780}{\nano \meter}, is still negligible and the Strehl ratio indicates diffraction limited performance.

\begin{table}[t]
\begin{center}
    \caption{Predicted performance of at system of two high-NA lenses in vacuum, for three wavelengths of interest.  In each case, the divergence $\Theta$ is chosen to maximize $S_1$, the Strehl ratio after one lens. $w_L(z_L)$ is the beam waist at the lens. Negative values of $\Theta$ indicate that the beam is convergent at the lens input. We also report $S_2$, the Strehl ratio after two lenses and $w_0$, the beam waist at focus.}
    \begin{tabular}{ c  c  c  c  c  c }
    \hline
   $ \lambda$ & $w_L(z_L)$  &  $\Theta$   &  $S_1$ & $S_2$ & $w_0$ \\ \hline
    \SI{780}{\nano \meter}     & \SI{2.2}{\milli \meter}  & -1.2 mrad   & 0.91   & 0.78  &  \SI{0.92}{\micro \meter}   \\ \hline
     \SI{795}{\nano \meter}      & \SI{2.2}{\milli \meter}  & -1.3 mrad   & 0.92   & 0.8    &  \SI{0.97}{\micro \meter}    \\ \hline
     \SI{852}{\nano \meter}     & \SI{2.2}{\milli \meter}  & -1.7 mrad   & 0.94   &0.86   &  \SI{1.03}{\micro \meter}    \\
    \hline
    \end{tabular}
    \label{tab1}
\end{center}
 \end{table} 

\section{Materials and techniques}
\label{sec:MaterialsAndTechniques}
Precise lens placement was accomplished using a three axis stage (Thorlabs MBT616D/M) with sub-\SI{}{\micro \meter} resolution plus two additional degrees of freedom of a pitch and yaw platform (Thorlabs PY003/M).
The lens holder, attached to this five-axis positioner, employed a pincer design to grab the lens by its perimeter, which protrudes beyond the lens' optical aperture. This holding method symmetrically distributes the forces to minimize stress-induced birefringence and aberrations.

\begin{figure}[ht]
\centering
\includegraphics[width=1\columnwidth]{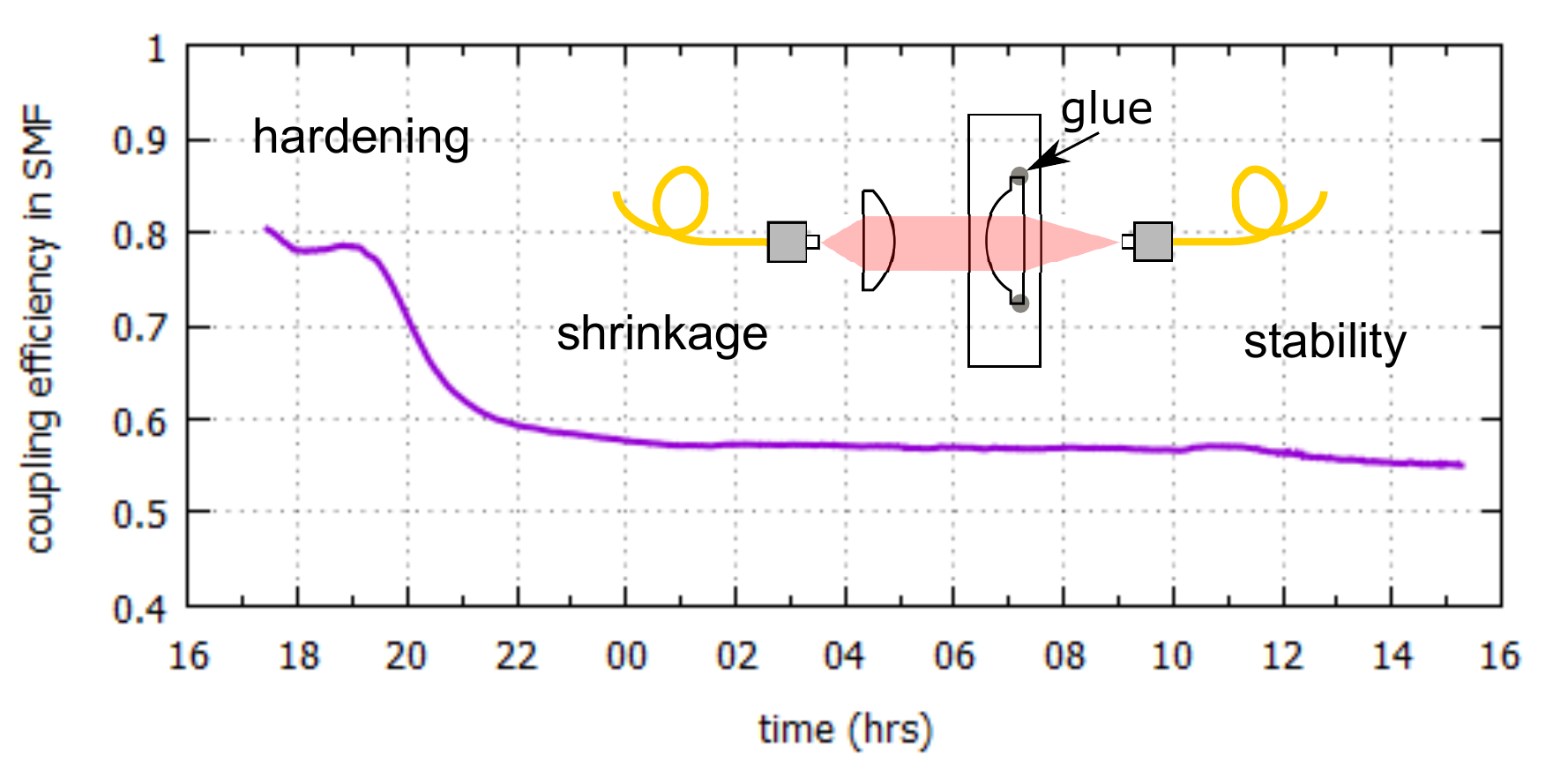}
\includegraphics[width=0.9\columnwidth]{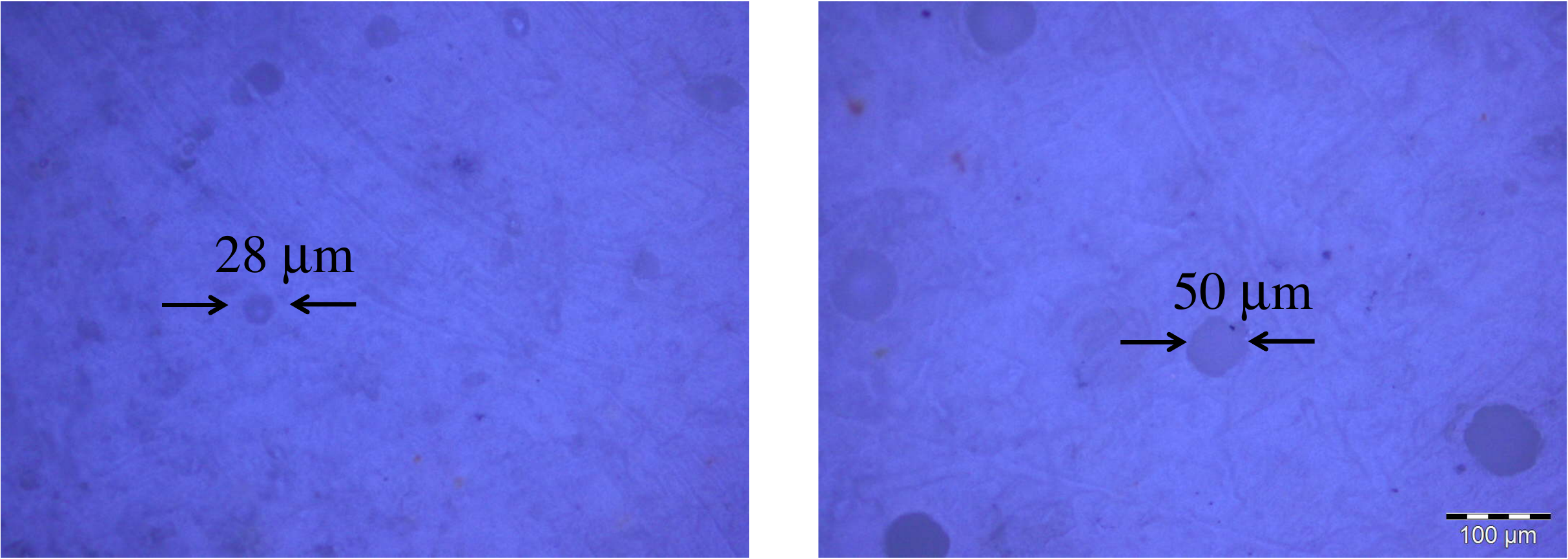}
\caption{ Left: Setup and measurement of the stability of the coupling efficiency of light into a SMF through a lens glued with Torr Seal as a function of the curing time of the epoxy. Right: reflection optical micrographs of two samples of glue after curing, the sample on the right was kept at \SI{10^{-3}}{m bar} for 20 minutes, in order to expand the residual bubbles and pull them near the surface. }
\label{gluing}
\end{figure}

After positioning (as described below), the lenses were glued using an ultra-low outgassing two-component epoxy (Varian Torr Seal) to an annular base made of the machinable ceramic Macor. Macor was chosen for its small coefficient of thermal expansion (\SI{9.3e-6}{\kelvin ^{-1}}) and low outgassing. 
As per the supplier's recommendations, we deposit the epoxy using a syringe with a narrow-bore needle to minimize the formation of air bubbles, and we maintain the assembly of lenses, base, and positioners at \SI{10^{-3}}{\milli bar} for the first 20 minutes of the cure, to allow bubbles to expand and move toward the surface. Fig.\ \ref{gluing} (bottom) shows the expansion of the air bubbles resulting from this procedure. The assembly is then taken out of vacuum, the lens is precisely positioned using the procedures described in Sec.~\ref{sec:AlignmentStrategy}, and allowed to cure for 24 hours before the lens is unclamped from the positioner.   

The nominal linear shrinkage of Torr Seal during curing is \SI{1.25e-3}{}, or \SI{1.25}{\micro\meter} for a \SI{1}{\milli\meter} thick bonding layer.  To measure the effect of shrinkage, we first coupled a gaussian beam into a single-mode fibre via an aspheric lens. We then applied the gluing procedure described above, and monitored the fibre in-coupling efficiency as a measure of the lens displacement due to shrinkage. As shown in Fig.\ \ref{gluing} (top), after two hours the shrinkage of the epoxy can be observed as a decrease of the coupling efficiency from 0.8 to 0.6 in a time window of about two hours, after which the system reaches stability. This loss of coupling corresponds to a displacement of the focused spot by less than the core diameter of the fiber (\SI{5}{\micro \meter}) in the transverse plane or less than the Rayleigh range (\SI{\sim 25}{\micro \meter}) in the longitudinal direction.

\section{Alignment procedure}
\label{sec:AlignmentStrategy}
\begin{figure}[ht]
\centering
\includegraphics[width=0.99\columnwidth]{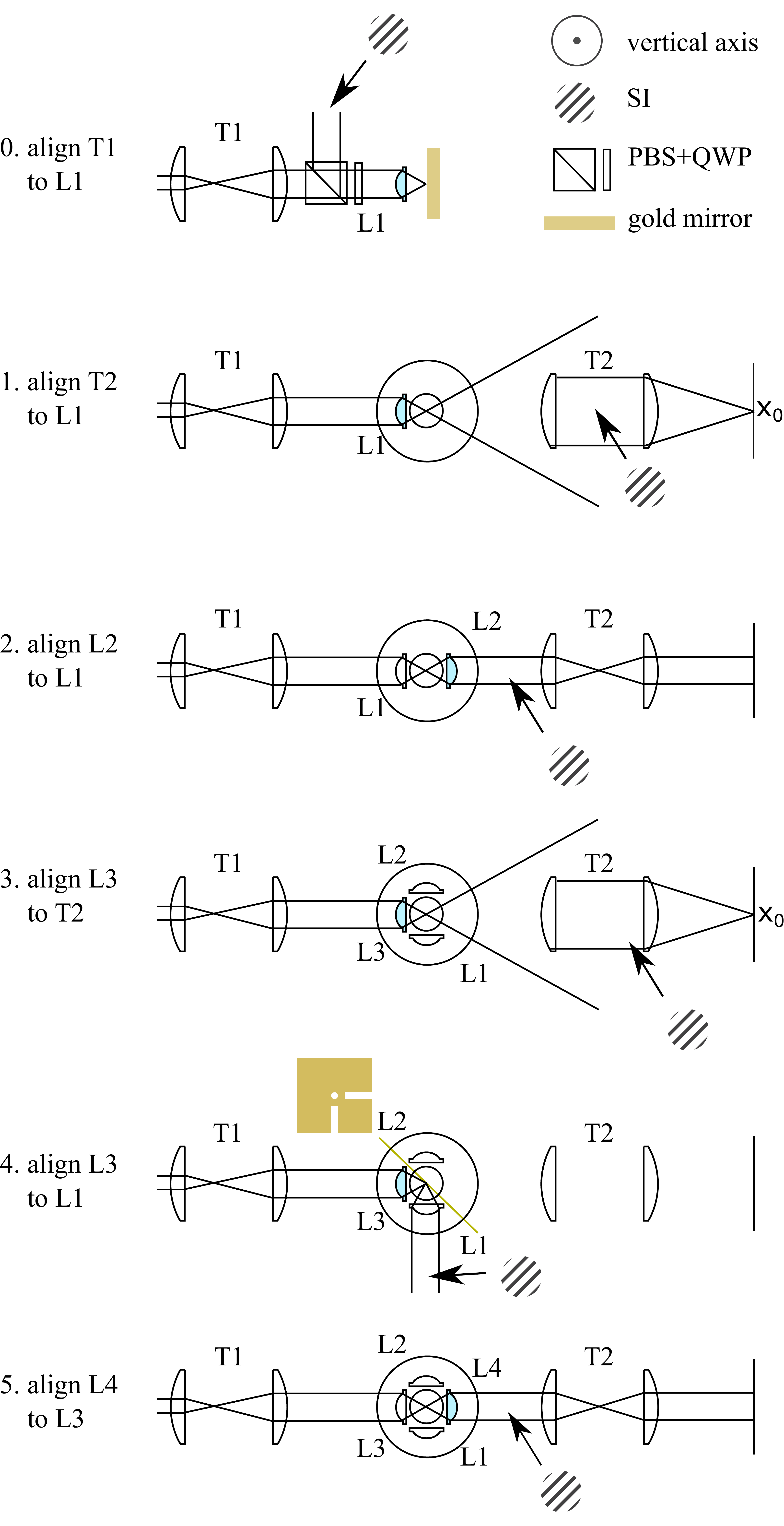}
\caption{ Procedure for positioning the lenses. See text for details.}
\label{procedure}
\end{figure}

The procedure for aligning the four lenses of the optical assembly is illustrated in Fig.~\ref{procedure}.  Throughout, an alignment laser at \SI{795}{\nano\meter} is used.  The steps (numbered in correspondence with the figure) are

\begin{enumerate}
\setcounter{enumi}{-1}
\item Telescope T1, with adjustable inter-lens separation, is used to produce a nearly-collimated beam of waist \SI{2.2}{\milli \meter} at the lens, which corresponds to a beam with ${\rm NA} = 0.26$ at \SI{795}{\nano\meter}, with adjustable divergence.  A polarizing beamsplitter followed by a quarter waveplate is used to sample the retro-reflected beam, which is analyzed with a SI. Lens L1 is centred in the input beam and adjusted to normal incidence by observation of the (weak) back-reflection from the lens itself.  A gold-coated first-surface mirror is placed at lens focus. The mirror tilt is adjusted for retro-reflection of the beam, and the mirror axial position is adjusted to minimize aberrations seen on the SI, which is simultaneously used to measure and adjust the divergence.  T1 is then fixed for the remainder of the alignment procedure. 
\item The ceramic support is now added, supported by a rotary stage about the vertical axis. This is followed by a 1:1 telescope {T2} to collect the output of L1 and focus it onto the sensitive surface of a CMOS camera with a pixel size of \SI{5.2}{\micro \meter}. The spot occupies $\approx \SI{3}{pixel}$ on the camera. The first lens of T2 is positioned to produce a collimated beam, as measured by a SI.  L1 is now glued in place as described in Sec.~\ref{sec:MaterialsAndTechniques}. Displacement of L1 during curing would be detectable as a displacement of the focused spot on the CMOS camera, with resolution $\sim\SI{5.2}{\micro \meter}$. In practice, we did not observe any displacement within the resolution of this technique. \item Lens L2 is aligned using a SI to minimize aberrations and set the output divergence equal to the input convergence. L2 is glued in place as described in Sec.~\ref{sec:MaterialsAndTechniques}. T2 is not used for this step.
\item The ceramic support is rotated by \SI{90}{\arcdeg} using the rotary stage. Lens L3 is now added and aligned, using back-reflection from the lens itself as with L1.  The longitudinal position of L3 is adjusted to minimize the spot size on  the CMOS camera after T2. The SI between lenses of T2 provides a check that the beam is collimated at this point. We estimate the precision of this procedure for setting the longitudinal position of L3 is \SI{\pm 100}{\micro \meter}.
\item We use a custom-coated gold first-surface mirror (described below), introduced at \SI{45}{\arcdeg} relative to the L1-L2 axis, to reflect the beam focused by L3 toward L1.  To position the mirror at the focus of L3, we make use of a \SI{1}{\micro\meter}-wide uncoated stripe on the mirror, which when positioned at focus transmits the beam to T2 and the CMOS camera without visible diffraction.  The mirror is then translated parallel to its surface by a few \SI{}{\micro\meter} until the beam is fully reflected.  A SI after L1 is used to measure the resulting collimation and aberrations. We estimate the precision of this procedure for setting the longitudinal position of L3 is \SI{\pm 25}{\micro \meter}. L3 is glued after this mirror-based adjustment.
\item  Lens L4 is placed and glued using the same procedure as for L2. 
\end{enumerate}

We note that a system of four aspheres around a central point have been assembled previously using somewhat different methods \cite{MartinezDorantesThesisBonn2016, JoseGallegoThesisBonn2017, GallegoPRL2018}.  In the cited works, the authors first align two aspheres at a right angle using a small reflective sphere as a reference. The aspheres are positioned to produce beams that retro-reflect from the sphere's surface, thereby guaranteeing mutual focus at the sphere center.  A shearing interferometer is used to measure and minimize aberrations in the retro-reflected beams. Two more aspheres are then aligned to the ones already placed, in the same manner as steps 2 and 5 above.  This procedure should, provided the sphere is ideal, yield the same precision as our technique.  We note that these publications do not describe simultaneous coupling to a single atom, nor do they describe high-NA measurements of the mutual overlap of the beam, leaving open the question of whether the method in fact succeeded in producing diffraction-limited coupling of the four lenses to a single point.

\section{Post-assembly characterization and in-vacuum alignment}
\begin{figure}[ht]
\centering
\includegraphics[width=0.90\columnwidth]{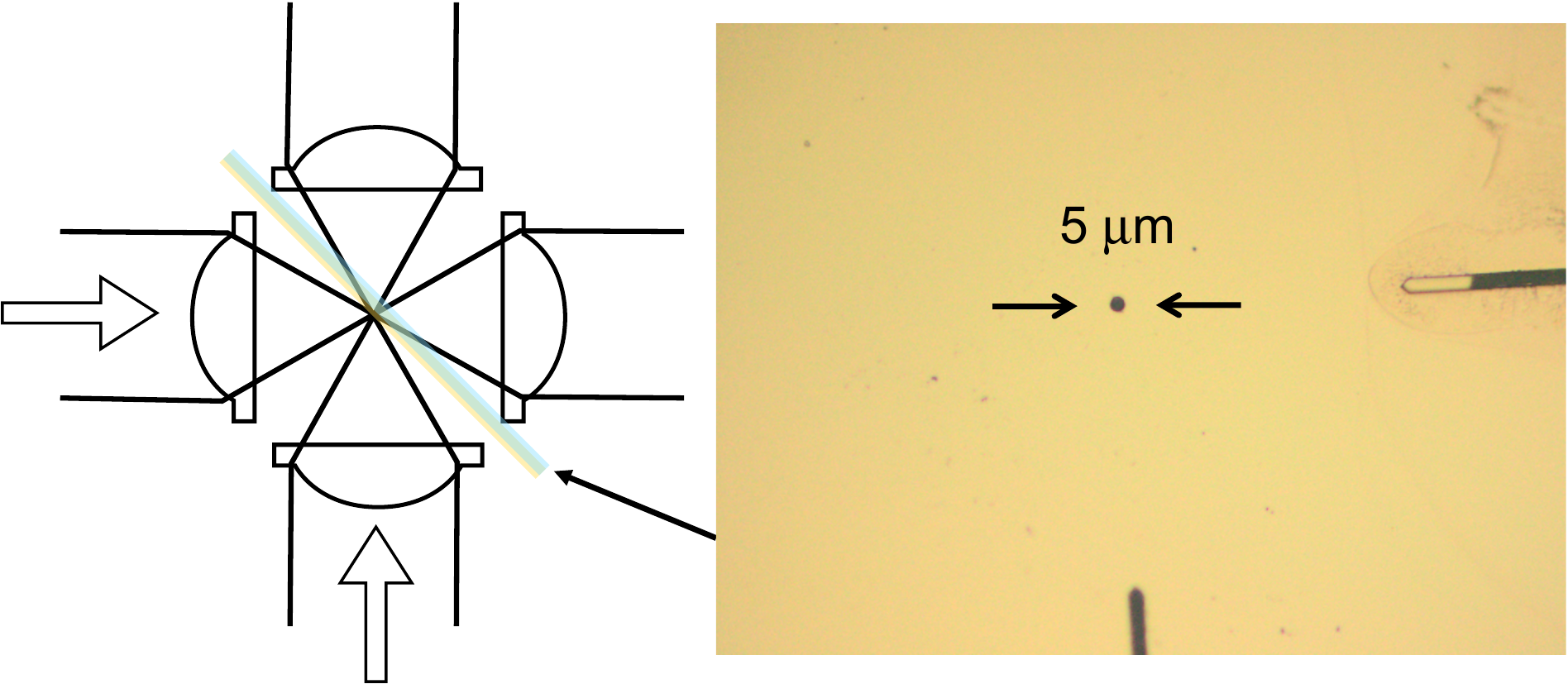}
\caption{High-NA focus localization using a micro-fabricated mirror with transmissive aperture.  Left:  illustration of the geometry showing four lenses and a gold-coated cover slip (green, indicated by arrow). Cover slip position, controlled by a micro-positioner, is used to locate the beam foci.   Right: Reflection optical micrograph of the mirror with the \SI{5}{\micro \meter} aperture, made with optical lithography on a gold-coated quartz plate. }
\label{mirror}
\end{figure}

If the above  lens assembly procedure succeeded, the four lenses should be able to form diffraction-limited images of the same single point in space.  Equivalently, it should be possible to pass a beam through the L1-L2 pair without introducing aberrations, and similarly through the L3-L4 pair, while also having these beams reach focus at the same point in space. We note that for this objective it is sufficient for the four lenses' diffraction-limited fields-of-view (DLFoVs) to share a non-zero overlap.  

We test this latter condition with the aid of a second gold-coated quartz cover slip, with a \SI{5}{\micro \meter} diameter circular aperture on the coated surface (see Fig.\ \ref{mirror}). In a first measurement, and for preliminary alignment of the measurement to follow, a beam with nominal divergence is passed through each lens pair, and aligned to minimize aberrations at the output as measured by SI.  The cover slip is then introduced into the focal region at \SI{45}{\arcdeg} and used to localize the two beam foci in 3D.  

\begin{figure}[ht]
\centering
\includegraphics[width=0.95\columnwidth]{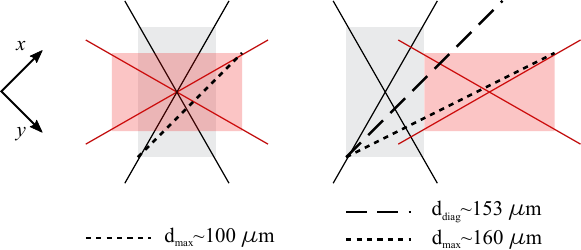}
\caption{Geometry of diffraction-limited fields of view (DLFoVs) for different lens positioning. The DLFoVs of the two pairs of lenses are represented by a grey and a red area for the vertical and horizontal pair, respectively. $x$ and $y$ indicate the directions of aperture translation, normal and in-plane of the mirror.  Left: DLFoVs for ideally-positioned lenses.  Foci can be separated (along $x$ or $y$) by up to  \SI{100}{\micro \meter} while remaining aberration-free.   Right: Closest-approach non-overlapping DLFoVs.  Foci can be separated (along $x$ or $y$) by up to  \SI{153}{\micro \meter} while remaining aberration-free. }
\label{fig:fig6}
\end{figure}

If the lenses are ideally located, this procedure is expected to position the foci near each other, with a precision comparable to the extent of the DLFoV.  As shown in Fig.~\ref{fig:fig6}, in the case of ideal lens positioning, and thus co-centric DLFoVs, the maximum separation would be $\approx \SI{100}{\micro \meter}$ along either diagonal direction $\left(x,y\right)$. If the DLFoVs do not overlap, the largest separation between the foci (again along the $x,y$ directions) is at least \SI{153}{\micro \meter}. In ten repetitions of the alignment procedure, the foci were always found to be separated by less than \SI{100}{\micro \meter} along the $\left( x,y\right)$ directions, suggesting that the DLFoVs shared a significant overlap.

Finally, we superimpose the foci, by positioning the micro-aperture within the focal region and adjusting the angle and divergence of the two input beams such that both pass through the aperture.  After this procedure, no aberrations are visible in the transmitted beams. This confirms that the procedure succeeded in overlapping the four lenses' DLFoVs, by directly showing simultaneous, diffraction-limited focus of the four lenses at a single point. 

The lenses were then placed in a UHV chamber suitable for atom trapping and cooling experiments.  Each lens is accessible via a viewport normal to the lens axis. Gaussian beams are produced using fibre collimators (Schafter and Kirchoff model 60FC-4-A15-02) and low-NA beam-expansion telescopes. For each lens pair, one beam is first adjusted to the nominal divergence using a SI, and then sent through the lens pair.  Alignment and focusing are adjusted to minimize aberrations as seen on a SI at the output. 

\begin{figure}[ht]
\centering
\includegraphics[width=0.95\columnwidth]{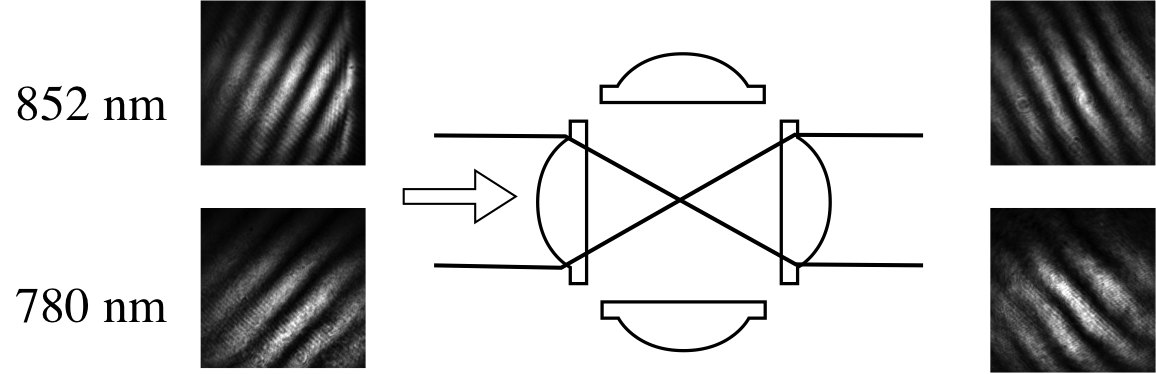}
\caption{ Shearing interference fringes for  \SI{780}{\nano\meter} (bottom, $\Theta_{780} \simeq$ \SI{1.2}{m rad}) and  \SI{852}{\nano\meter} (top, $\Theta_{852} \simeq $ \SI{1.7}{m rad}). The images were taken with lenses in vacuum and with the foci of the four \SI{780}{\nano\meter} collection beams overlapped with the focus of the \SI{852}{\nano\meter} optical dipole trapping beam. The equal and opposite fringe tilts indicate reflection symmetry about the centre, and thus focal overlap at this point.  }
\label{Shearing}
\end{figure}

As a final test of the diffraction-limited performance of the lens system, we again use the SI to check for aberrations at \SI{780}{\nano\meter} in each pair of collinear lenses, and at \SI{852}{\nano\meter} for the pair of lenses creating the dipole trap.  We note that at this point the foci of the four lenses are overlapped, both at \SI{780}{\nano\meter} and at \SI{852}{\nano\meter}, as evidenced by the single atom signals. 
Fig.\ \ref{Shearing} shows the measured fringes for \SI{780}{\nano\meter} and  \SI{852}{\nano\meter}, which confirm the diffraction-limited focusing in this condition.

We now couple fluorescence from a dipole-trapped atom into Single Mode Fiber (SMF) via the trap-axis lenses L1 and L2. This is facilitated by overlapping an auxiliary beam emitted from the collection fiber with the trapping beam. To find a similar signal with the right-angle lenses L3 and L4, rubidium vapour is introduced into the UHV chamber and resonance fluorescence, imaged on a CMOS camera, is used to overlap the beam foci as shown in Fig.~\ref{fig:Overview} (upper right). At this point we are able to directly image a single dipole-trapped atom held in the FORT through any of the lenses.  Fine alignment of a single-mode fibre behind L4 to the trapped atom is facilitated by sending a weak beam from the fibre and imaging both the single trapped atom and the introduced beam through L3.
 
 \section{Single-atom signals}
We continuously run a MOT, including cooler and repumper beams at \SI{780}{\nano\meter}, and a single-beam FORT at \SI{852}{\nano\meter} strongly focused through L1. We collect atomic fluorescence into single-mode fibres behind each of the four lenses and detect with single-photon-sensitive avalanche photodiodes.  Fig.\ \ref{fig:Overview} (bottom left) shows the observed signals, which show a characteristic random telegraph signal alternating between no atom with a low photon count due mostly to background MOT fluorescence, and one atom with a higher photon count, due to fluorescence excited by the MOT beams.  For lens L1, these levels are \SI{\sim 1.5e3}{counts/s} and \SI{\sim 9e4}{counts/s}, respectively, and permit a statistically strong discrimination of one atom from no atom in under \SI{10}{\milli\second}. 

Averaging the count rate in the intervals marked in green, and subtracting the mean count rate in the regions marked in blue, we compute the mean 1-atom contribution to any given channel's count rate.  We then compute the efficiency ratio (L1:L2:L3:L4) to find 1:0.99:0.42:0.37. The result is consistent with the expected collection-efficiency ratios \cite{TeyNJP2009} assuming diffraction-limited collection and an atomic temperature of \SI{120}{\micro\kelvin}, which is typical for single rubidium atoms in strongly-focused dipole traps \cite{FuhrmanekNJP2010}. The difference reflects the fact that the trapping potential and atomic probability distribution are elongated along the trap axis, while the collection efficiency of any given lens is more tolerant to longitudinal displacements of the source than to transverse ones. These observations confirm what is seen in the wavefront measurements, namely that the coupling is simultaneously diffraction-limited from all four directions.

We analyze right-angle coincidence detection events between the L1 and L3/L4 channels, limited to intervals in which an atom is observed, to measure the autocorrelation function $g^{(2)}(\tau)$.  The result, shown in Fig.\ \ref{fig:Overview} (bottom), shows oscillations at the generalized Rabi frequency and a minimum at $\tau=0$ of  $g^{(2)}(0)=0.44\pm0.06$,compatible with the background level (in grey). A value below unity rigorously shows the non-classical nature of the light emitted from the trap \cite{Kimble:1977aa}, \cite{MandelOL1979} and confirms the presence of a single quantum emitter. The residual value of $g^{(2)}(0)$ is due to the scattered MOT light, which gives a background of Poisson-distributed events.

\section{Conclusion}
We have assembled four high-NA aspheric lenses in a Maltese cross geometry with a common, diffraction-limited central focus. By trapping a single atom at this focus and collecting anti-bunched atomic fluorescence from it, we have demonstrated compatibility of this optical technology with cold-atom and quantum optical techniques. The use of four lenses immediately doubles the available solid angle relative to the corresponding two-lens geometry, to give a large coupling boost for quantum optical and quantum technological applications profiting from large solid angles. The geometry can be used to make wavelength-scale and sub-wavelength potentials using right-angle dipole traps and or optical lattices. Right-angle access will also enable the study of new processes, e.g. sub- and super-radiance at large angles. The same strategy could be applied for large solid-angle coupling to ions, molecules, nano-spheres, nano-diamonds, and other species that can be optically manipulated in free space. 

\section*{Funding Information}
This project was supported by the European Research Council (ERC) projects 
AQUMET (280169) and  ERIDIAN (713682); European Union projects QUIC (Grant Agreement no. 641122), FET Innovation Launchpad UVALITH (800901), Quantum Technology Flagship project MACQSIMAL (820393) and QRANGE (820405); EMPIR project USOQS (17FUN03); the Spanish MINECO projects MAQRO (Ref. FIS2015-68039-P) and Q-CLOCKS (PCI2018-092973), XPLICA  (FIS2014-62181-EXP), the Severo Ochoa programme (SEV-2015-0522); Ag\`{e}ncia de Gesti\'{o} d'Ajuts Universitaris i de Recerca (AGAUR) project (2017-SGR-1354); Fundaci\'{o} Privada Cellex, Fundacio Privada MIR-PUIG,  and Generalitat de Catalunya (CERCA) program.

\section*{Acknowledgments}

The Authors thank Johann Osmond for the fabrication of the gold-coated quartz plates used for the alignment, Wenjamin Rosenfeld, Yvan Sortais, Antoine Browaeys, Thierry Lahaye, Ludovic Brossard and Ilja Gerhard for helpful discussions.

\bibliography{lensesbibshort}

\end{document}